\documentclass[3p,times,twocolumn]{elsarticle}

\usepackage{ecrc}

\volume{00}

\firstpage{1}

\journalname{Nuclear Physics B Proceedings Supplement}

\runauth{}

\jid{nuphbp}

\jnltitlelogo{Nuclear Physics B Proceedings Supplement}

\usepackage{amssymb}

\usepackage[figuresright]{rotating}

\newcommand{\lapprox}{\mathrel{%
\setbox0=\hbox{$<$}\raise0.6ex\copy0\kern-\wd0\lower0.65ex\hbox{$\sim$}}}
\newcommand\TT{\rule{0pt}{2.5ex}}        
\newcommand\BB{\rule[-1.0ex]{0pt}{0pt}}  
\newcommand{\be}{\begin{equation}}
\newcommand{\en}{\end{equation}}
\newcommand{\bes}{\begin{equation*}}
\newcommand{\ens}{\end{equation*}}
\newcommand{\bc}{\begin{center}}
\newcommand{\ec}{\end{center}}
\newcommand{\bt}{\begin{tabular}}
\newcommand{\et}{\end{tabular}}
\newcommand{\bg}{\begin{minipage}}
\newcommand{\eg}{\end{minipage}}
\newcommand{\ba}{\begin{array}}
\newcommand{\ea}{\end{array}}
\newcommand{\bea}{\begin{eqnarray}}
\newcommand{\ena}{\nonumber\end{eqnarray}}
\newcommand{\lbl}[1]{\label{eq:#1}}
\newcommand{\rf}[1]{(\ref{eq:#1})}
\newcommand{\lbltab}[1]{\label{tab:#1}}
\newcommand{\Table}[1]{\ref{tab:#1}}

\newcommand{\fig}[1]{\ref{fig:#1}}
\newcommand{\braque}[1]{{\langle #1 \rangle}}
\newcommand{\im}{\hbox{Im}}

\newcommand{\metad} {m^2_\eta}
\newcommand{\mpi}{m_\pi}
\newcommand{\meta}{{m_\eta} }
\newcommand{\mpid}{m_\pi^2}
\newcommand{\mkd}{m_K^2}

\newcommand{\Vmud}{\bar{u}\gamma^\mu{d}}
\newcommand{\ubard}{\bar{u}{d}}
\newcommand{\piplus}{{\pi^+} }
\newcommand{\fplus}{f^{\eta\pi}_+}
\newcommand{\fminus}{f^{\eta\pi}_-}

\newcommand{\fzero}{f^{\eta\pi}_0}
\newcommand{\kzero}{{K^0}}
\newcommand{\kplus}{{K^+}}
\newcommand{\pizero}{{\pi^0} }
\newcommand{\disc}{\hbox{disc}}
\newcommand{\bleu}[1]{#1}
\newcommand{\rouge}[1]{#1}

\newcommand{\negatspace}{\!\!\!\!\!\!\!\!\!\!\!\!\!\!\!\!}

\begin{document}

\begin{frontmatter}



\dochead{}

\title{Dispersive evaluation of the second-class amplitude
  $\tau\to\eta\pi\nu_\tau$ \\ in the standard model}


\author[label1]{S. Descotes-Genon}
\author[label2]{E. Kou}
\author[label3]{B. Moussallam}

\address[label1]{LPT, Universit\'e Paris-Sud 11, Orsay}
\address[label2]{LAL, Universit\'e Paris-Sud 11, Orsay}
\address[label3]{IPN, Groupe Th\'eorie, Universit\'e Paris-Sud 11,
  Orsay}

\begin{abstract}
We reevaluate the two form factors relevant for the $\eta\pi$
second-class $\tau$ decay mode, making systematic use of analyticity, 
unitarity, combined with 
updated inputs to the NLO chiral constraints. 
We focus in particular on the shape of the $\rho$ resonance peak which
is a background-free signature of a second-class current. Its
dispersive construction requires the $\eta\pi\to\pi\pi$ scattering
amplitude which we derive from a family of Khuri-Treiman equations
solutions constrained with accurate recent  results on the $\eta\to3\pi$
Dalitz plot. 
\end{abstract}


\end{frontmatter}


\section{Introduction}
\label{Introduction}

Weinberg remarked that exact isospin conservation by the
strong interactions would imply selection rules for semi-leptonic weak
decays~\cite{Weinberg:1958ut}. Isospin conservation, of course, is
only approximate, being broken both in QCD because $m_u\ne m_d$ and in
QED because $q_u\ne q_d$. Yet, no quantitative experimental evidence
for these so-called second-class currents have been reported to
date. Processes of this type in $\tau$ decays are the $\eta\pi$ or
$\eta'\pi$ modes: because of parity conservation these decays must
proceed through the $I=1$ vector current which is even under a
$G$-parity rotation, while $\eta\pi$, $\eta'\pi$  are odd eigenstates
of $G$-parity. The sensitivity of these modes to physics beyond the
standard model (SM) has been discussed in
refs~\cite{Bramon:1987zb,Nussinov:2008gx}. Within the standard model,
measuring these $\tau$ decay amplitudes would provide non trivial
informations on matrix elements of the scalar operator
$\partial_\mu\Vmud$ and a related determination of the $\bar{q}q$
content of the scalar  $a_0(980)$, $a_0(1450)$ resonances. The Babar
collaboration has published upper bounds for the $\eta\pi$, $\eta'\pi$
modes~\cite{delAmoSanchez:2010pc,Aubert:2008nj}.  

We reconsider here the theoretical expectation for the
$\tau\to\eta\pi\nu$ mode in the SM.  This problem was first addressed
in ref.~\cite{Tisserant:1982fc}. We attempt to refine the theoretical
prediction by exploiting the  analyticity properties of the two form
factors involved and combining them with chiral symmetry
constraints. This proves particularly fruitful for the vector form
factor, for which unitarity provides a simple relation with the
isospin violating $\eta\to3\pi$ decay amplitude. We will show how
considerable recent progess in measuring this amplitude impacts the
determination of the $\eta\pi$ vector form factor.  It is usually
expected that the integrated branching fraction should be dominated by
the scalar rather than the vector form factor. However,
the main experimental obstacle to the observation of second-class
amplitudes at B-factories is  the pollution from first-class
background contributions (e.g. from $\tau^\pm\to \eta\pi^\pm\pi^0\nu$,
$\eta\pi^\pm K^0\nu$ where the extra neutral particle escapes
detection). While a peak in the $\eta\pi$ invariant mass at the
$a_0(980)$ mass may  be present in background modes, a peak at the
$\rho(770)$ mass unmistakingly signals a second-class contribution.

\section{Dispersion relations and chiral symmetry}
\label{Dispersion}
The $\eta\pi$ form factors satisfy simple analyticity properties, in
exactly the same way as the more familiar $\pi\pi$ or $K\pi$ form
factors. They can be defined as  analytic functions of the energy
variable, $s=(p_\eta+p_\pi)^2$ with a right-hand cut and, furthermore,
they are expected to behave as $1/s\log(s)$ when
$|s|\to\infty$. Consequently, they satisfy unsubtracted dispersion
relations (DR's). In practice, it is judicious to write DR's
for the form factors multiplied by weight functions like
$1/s^n$. For instance, if we consider $f_+^{\eta\pi}(s)/s^2$, the DR
writes 
\be\lbl{drfplus}
\!\!\!\!\!\!\!f_+^{\eta\pi}(s)= f_+^{\eta\pi}(0)+s \dot{f}_+^{\eta\pi}(0)+
{s^2\over\pi}\int_{4\mpid}^\infty ds' {\disc[ f_+^{\eta\pi}(s')]\over
(s')^2 (s'-s) }\ 
\en 
Thanks to the cutoff function, the integrand is dominated by the
energy region below 1 GeV where we can evaluate the discontinuity,
using unitarity, with only a few channels contributing (essentially,
only a single channel). The price to pay is that we have to provide
the values of the form factor and its derivative at $s=0$. For this
purpose, we can rely on three flavour chiral symmetry, since the
$\pi$ and the $\eta$ are both pseudo-Nambu-Goldstone bosons in this
framework.  

The vector and scalar $\eta\pi$ form factors are defined
starting from the matrix element of the vector current
$$
\langle\eta\pi^+\vert\bleu{\Vmud}\vert 0\rangle\!\!=\!\!-\sqrt2\left[
\bleu{\fplus(s)}\,(p_\eta-p_\pi)^\mu
\!+\!\bleu{\fminus(s)}\,(p_\eta+p_\pi)^\mu\right]\nonumber
$$
and 
\be\lbl{scalardef}
\!\!\!\!\!\!\!\bleu{\fzero}(s)
=\bleu{\fplus}(s)+{\displaystyle  s\over\displaystyle \Delta_{\eta\pi}}
\bleu{\fminus}(s)\ ,\quad \Delta_{\eta\pi}=\metad-\mpid
\en
At LO in the chiral expansion, the two form factors are constant
and equal, 
$$
\left.\fplus(s)= \fzero(s)\right\vert_{LO} 
= \rouge{\epsilon}={\displaystyle\sqrt3(\rouge{m_d-m_u})\over
  \displaystyle4\,\,(\rouge{m_s-{m_{ud}}})}\simeq 0.99\times10^{-2}\ .
$$
There is no electromagnetic contribution at this order and the
numerical estimate uses the chiral expansion of the mass differences
$m^2_{K^+}-m^2_{K^0}$, $m^2_{\pi^+}-m^2_{\pi^0}$ also at LO.

The  two form factors were computed at NLO in the chiral expansion  by
Neufeld and Rupertberger~\cite{Neufeld:1994eg}, including also the EM
contributions at order $e^2$.  A remarkably simple expression
emerges from their results, relating the value  of the
$\eta\pi$ form factors at  $s=0$ to the ratio of the $K^+\pi^0$ and
$K^0\pi^+$ form factors  
\be
f_+^{\eta\pi}(0)={1\over\sqrt3}\left[\frac{f_+^{\kplus\pizero}(0)}{
  f^{\kzero\piplus}_+(0)}-1-{3e^2\over4(4\pi)^2}
\log{\mkd\over\mpid}\right]\ 
\en
Exploiting the recent results on $K^+_{l3}$ and $K^0_{l3}$ decays from
$K$ factories (see e.g.~\cite{Antonelli10}) yields  the most precise
evaluation of the $\eta\pi$ form factors at $s=0$,
\be\lbl{fplusval0}
f_+^{\eta\pi}(0)=f_0^{\eta\pi}(0)=(1.49\pm0.23)\times10^{-2}
\en
which is significantly enhanced from its LO estimate.

\section{Vector form factor and $\eta\to3\pi$}
The discontinuity of $\fplus$ can be associated with a sum over
intermediate states of the matrix element of the vector current
\be\lbl{unitrel}
\im\braque{\eta\pi^+\vert\bar{u}\gamma^3{d}\vert0}={1\over2}\sum_n
T^*_{n\to\eta\pi^+}\braque{n\vert\bar{u}\gamma^3{d}\vert0}\ .
\en
The derivation is valid in the unphysical situation where the $\eta$
meson is stable and we will assume than an analytic continuation as a
function of $\meta$ is possible. Below 1 GeV, the sum in
eq.~\rf{unitrel} is essentially saturated by the contribution of the
lightest state $n=\pi^0\pi^+$ (as it is strongly enhanced by its coupling
to the $\rho$ resonance). This leads to the following estimate of the
discontinuity ( for $4\mpid \le s \lapprox 1$ GeV$^2$)
\bea\lbl{discfplus}
&&\disc[\fplus(s)]=-\theta(s-4\mpid)\times\\
&&{s-4\mpid\over32\pi\sqrt{\lambda_{\eta\pi}(s)}}
F_V^\pi(s)\int_{-1}^1 dz z T^*_{\pi\pi\to\eta\pi}(s,t(z))
\ena
with $\lambda_{\eta\pi}(s)=(s-m_-^2)(s-m_+^2)$, $m_\pm=\meta\pm\mpi$. In
this equation, $F_V^\pi$ is the pion vector form factor, which is 
precisely known experimentally. In addition, one needs to evaluate
the $\pi\pi\to \eta\pi$ amplitude projected on the $P$-wave,
partly in an unphysical region ($s<m_+^2$). It can
be determined  using its analyticity properties together with
experimental constraints on $\eta\to 3\pi$ decay.  

Combining analyticity with elastic unitarity for $\pi\pi$
(re)scattering leads to a system of Khuri-Treiman (KT)
equations~\cite{KT}. The solutions of these equations in their full
generality were first discussed in
refs.~\cite{Kambor:1995yc,Anisovich:1996tx}. A subtle point, in
particular, concerns the treatment of the singularities of the
partial wave projected $\pi\pi\to\eta\pi$ amplitude (
thus, $T^{J=1}_{\pi\pi\to\eta\pi}\sim 1/(s-m_-^2)^{3/2}$  when $s\to m_-^2$)
in the integrals. Eq.~\rf{discfplus} shows that these singularities
affect also the computation of the vector form factor and must be
treated by the same method. In practice, this leads to a distortion of
the shape of the $\rho$ resonance as compared to a naive vector meson
dominance (VMD) approach. 

The authors of ref.~\cite{Anisovich:1996tx}  argue that a
four-parameter family of solutions are relevant for the $\eta\to3\pi$
decay, which we can also use for our problem.  Assuming a sufficiently
fast convergence of the three-flavour chiral expansion, they propose
to determine all these four parameters by matching the dispersive and
NLO chiral amplitude in a region around the Adler zero.  
Unfortunately, the amplitude obtained in this manner turns out
not to be in agreement with the experimental results on the Dalitz
plot parameters (see table~\Table{dalitz} below). One must thus
determine the KT solution parameters partly from matching to the NLO
amplitude and partly from fitting the experimental Dalitz plot
data~\cite{Colangelo:2011zz}. We perform here a fit analogous to
ref.~\cite{Colangelo:2011zz} but constraining the four KT parameters
to be exactly real. From the NLO amplitude, we use the position of the
Adler zero but not the value of the amplitude slope at this
point. Thus, we do not attempt to determine the value of the quark
mass ratio from the $\eta$ decay rate as
in~\cite{Colangelo:2011zz} (see
also~\cite{Bijnens:2007pr,Kampf:2011wr}) but take this ratio from the  
PDG (which leads to a value of the slope at the Adler  zero differing
from the NLO prediction by approximately 20\%).
\begin{table}[ht]
\begin{tabular}{@{}c|c@{}cc@{}}\hline
\TT param. & experimental  & NLO Match. & Fit \\ \hline
\TT a &$-1.090\pm0.005^{+0.008}_{-0.019}$ & $-1.300 $  &  $-1.065$   \\
 b &$ 0.124\pm0.006\pm 0.010    $ & $ 0.463 $  &  $0.159$  \\
 d &$ 0.057\pm0.006^{+0.007}_{-0.016}$ & $ 0.069$   &   $0.066$   \\
 f &$ 0.14 \pm0.01 \pm 0.02$      & $ 0.001$   &   $0.107 $
 \\ \hline\hline
\TT $\alpha$ & $-0.0315\pm 0.0015$ & $0.015$ &   $-0.0355$  \\ \hline
\end{tabular}
\caption{Comparison of the $\eta\to3\pi$ Dalitz plot parameters
  obtained from KT solutions with experiment. The Dalitz parameters
  $a$, $b$, $d$, $f$ refer to the charged decay mode and are taken
  from~\cite{kloe08}, while $\alpha$ refers to the neutral mode and the
  quoted value is taken from the PDG.}
\lbltab{dalitz}
\end{table}

\begin{figure}[htb]
\includegraphics[width=1.05\linewidth]{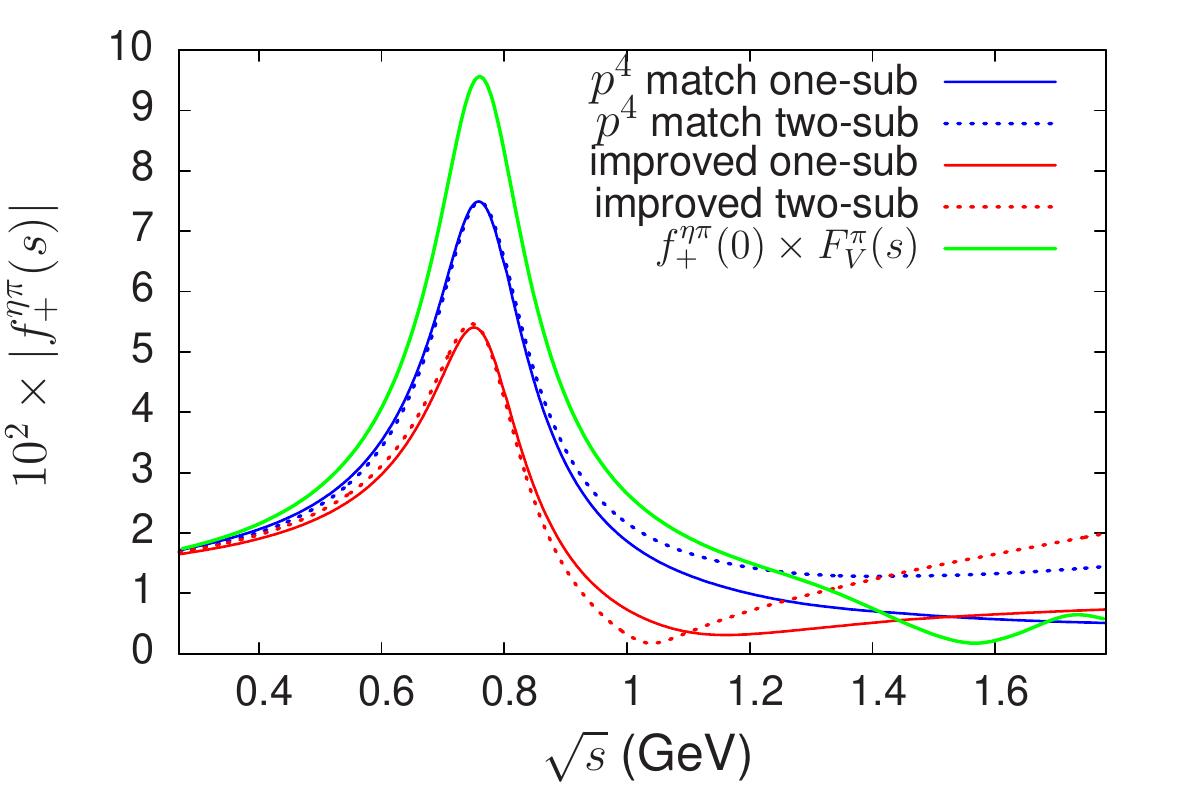}
\caption{\sl Comparison of a naive VMD model for $\fplus$ (upper
  curve) with dispersive calculations using KT solutions.}
\label{fig:fplus}
\end{figure}

The results of using a KT solution in the discontinuity
relation~\rf{discfplus} and then computing the form factor from the DR
with weight functions $1/s$ and $1/s^2$ and using NLO chiral
constraints like~\rf{fplusval0} is illustrated in
fig.~\fig{fplus}. The stability with respect to the weight functions
is satisfactory below 1 GeV. The results are also compared with the
naive VMD where $\fplus$ would be simply proportional to the
pion form factor normalized to the  value~\rf{fplusval0} at the
origin. The dispersive calculation is seen to yield a significantly
reduced resonance peak as compared to a naive VMD modelling. The
influence of the KT parameters is also rather significant and provide
an idea of the uncertainties of this calculation.

\section{Scalar form factor model}
\begin{figure}[htb]
\includegraphics[width=1.05\linewidth]{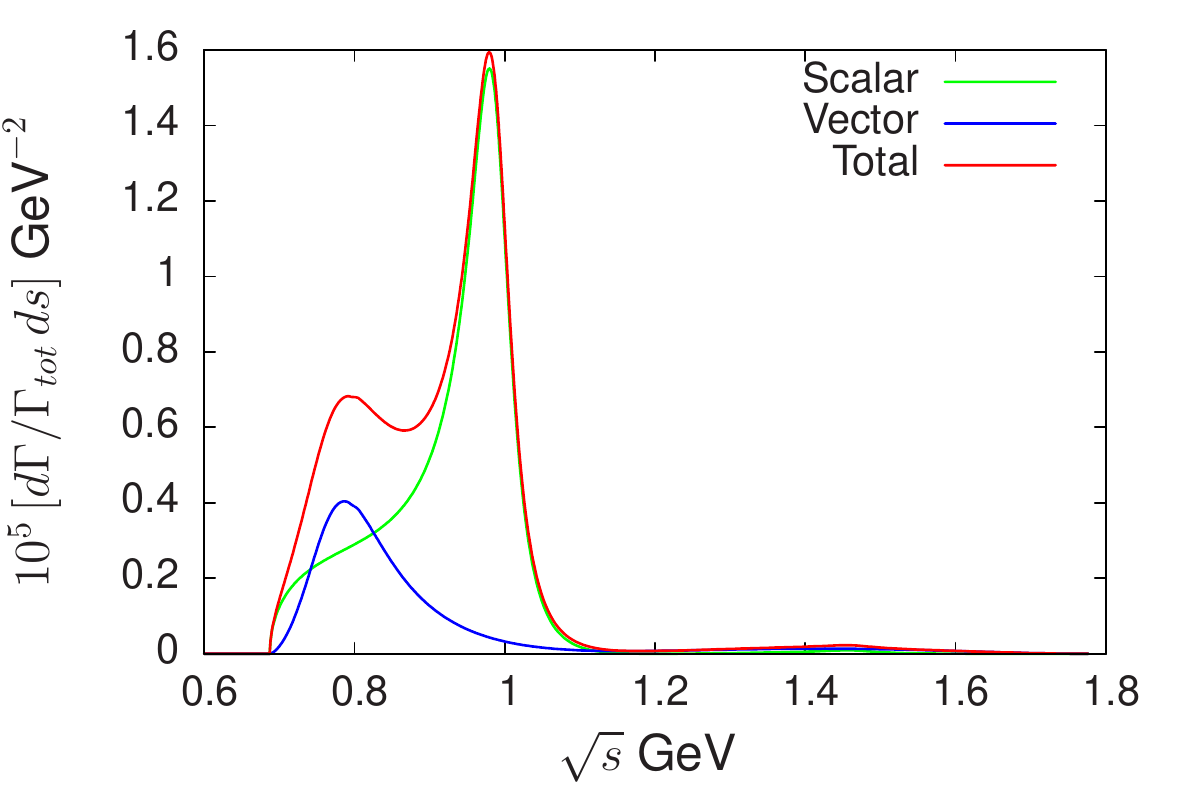}
\caption{Spectral function for $\tau\to \eta\pi\nu$ using central
  values for the dispersive constructions of the vector and the scalar
  $\eta\pi$ form factors.
} 
\label{fig:spectral}
\end{figure}

The scalar form factor (see~\rf{scalardef}) coincides with the
$\eta\pi$ matrix element of the derivative operator
$i\partial^\mu\Vmud$ and a  discontinuity relation analogous to
eq.~\rf{unitrel} can be written. Since $i\partial^\mu\Vmud$ is
itself isospin suppressed as can be seen from the Ward identity
$i{\partial_\mu\Vmud}= ({m_d-m_u}){\ubard} - {e} A_\mu {\Vmud}r$,
the sum over states
\footnote{In principle, one should include states $\vert
  n\gamma\rangle$ in the sum as required by the second term in the Ward
  identity. These contributions as well as other EM ones where the
  photon line attaches to the charged lepton are neglected here.} 
can be resticted to isospin conserving $T$-matrix elements
$T_{n\to\eta\pi}$. Below one GeV, the state $n=\eta\pi$ dominates and
we can write          
\bea\lbl{unitf0rel}
&&\negatspace \im\fzero(s)=\theta(s-m_+^2)\times\\
&&\negatspace \quad {\sqrt{\lambda_{\eta\pi}(s)}\over 16\pi{s}}
\fzero(s)\times{1\over2}\int_{-1}^1
dz\,T^*_{\eta\pi^+\to\eta\pi^+}(s,t(z)) 
\ena
For such a unitarity relation Watson's theorem implies that the phase
of the form factor coincides with the $\eta\pi$ scattering  phase shift
in the region of elastic scattering. It is then natural to employ a
phase dispersive representation for the form factor, e.g.
\bea\lbl{fzerodisp}
&&\negatspace \fzero(s)= \fzero(0)
\left({\fzero(\Delta_{\eta\pi}) \over \fzero(0)}\right)^{{s\over\Delta_{\eta\pi}}}\\
&& \negatspace \times\exp\left({\displaystyle s(s-\Delta_{\eta\pi})
\over\displaystyle\pi^{\phantom{'}}}{\displaystyle\int_{(\meta+\mpi)^2}^\infty}
 ds'{\displaystyle\phi^{\eta\pi}(s')
\over \displaystyle s'(s'-\Delta_{\eta\pi})(s'-s) }
\right) 
\ena
(which uses the weight function $1/s(s-\Delta_{\eta\pi})$
following~\cite{Bernard:2007cf}). The values of the form factor at
$s=0$ and the Dashen-Weinstein point $s=\Delta_{\eta\pi}$ must be
provided from NLO ChPT. A difficulty at this point is that the
$\eta\pi$ scattering phase shift is not measurable by the same methods
as used for $\pi\pi$ or $K\pi$. The experimental information concerns
the properties of the resonances which couple to $\eta\pi$ and the
phase shift is constrained near the threshold by chiral symmetry. We
will use a simple model proposed in ref.~\cite{schechteretapi99} which
interpolates between these pieces of information. This model makes the
plausible prediction that the global features of $\eta\pi$ scattering
\begin{figure}[th]
\includegraphics[width=1.05\linewidth]{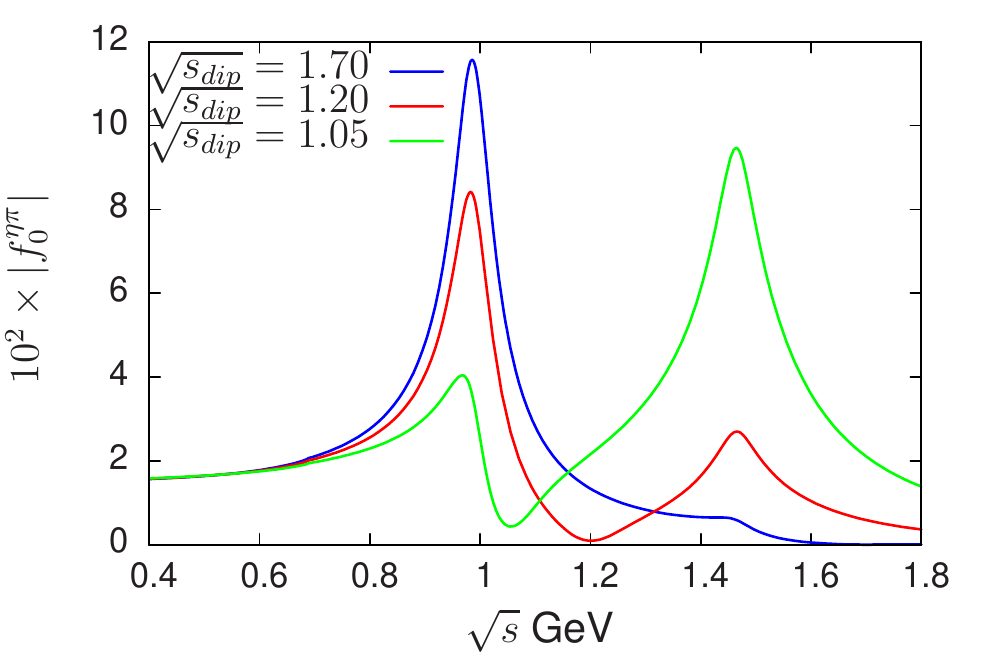}
\caption{\sl Scalar form factor modulus from the phase
  representation~\rf{fzerodisp} illustrating the effect of varying
  $s_{dip}$.}   
\label{fig:fzero_mod}
\end{figure}
are fairly similar to those of $\pi\pi$ and $\pi K$ scattering. The
phase shift is a steadily raising function and inelasticity sets in
rather sharply at a two-particle threshold ($K\bar{K}$) close to a
resonance ($a_0(980)$). We can make use of this analogy to make a
guess for the behaviour of the form factor phase in the inelastic
energy region. In the case of $\pi\pi$ or $\pi K$ the analogous phase
can be determined by solving a system of coupled Muskhelishvili-Omn\`es
equations   using a known set of $T$-matrix elements. A common feature
is that the phase drops  sharply by approximately $\pi$ close to the
inelastic threshold, which causes a dip in the modulus of the form
factor. The exact point $s_{dip}$ where this  happens in 
the case of $\eta\pi$ cannot, of course, be known without actually
solving the analogous equations, but it seems plausible that this
should be somewhere in between the two resonances $a_0(980)$ and
$a_0(1450)$. If one of these resonances can be interpreted as a
tetraquark state (i.e. having  a suppressed coupling to the $\ubard$
operator) then, from the analyticity point of vue, this corresponds to
$s_{dip}$ lying close to the corresponding
resonance. Fig.~\fig{fzero_mod} illustrates different values for
$s_{dip}$ and fig.~\fig{spectral} shows the complete spectral
function for $\tau\to \eta\pi\nu$ assuming $s_{dip}$ to be midway
between $a_0(980)$ and $a_0(1450)$.     

Finally, integrating over the spectral functions, we obtain the
branching fractions shown in the last line of
table~\Table{branchings}. The results are preliminary and the errors
quoted reflect only the uncertainty associated with the variation of
$s_{dip}$. Our results tend to be in the lower range of previous
evaluations. 

\begin{table}[ht]
\begin{tabular}{@{}llll|l}\hline\hline
\TT \ &  $10^5\times$BF$_V$ &  $10^5\times$BF$_S$   &  $10^5\times$BF & ref.   \\ \hline
\TT & 0.25 & 1.60 & 1.85 &  \cite{Tisserant:1982fc} \\
& 0.12 & 1.38 & 1.50 & \cite{Pich:1987qq}      \\
&  0.15 & 1.06 & 1.21 & \cite{Neufeld:1994eg}\\
&  0.36 & 1.00 & 1.36 & \cite{Nussinov:2008gx}\\
&  $[$0.2-0.6$]$ & [0.2-2.3] & [0.4-2.9]&  \cite{paverriazuddin10}\\ \hline
\TT \BB &  0.11 & $0.37^{+0.30}_{-0.20}$ & $0.48^{+0.30}_{-0.20}$ &
This work\\ \hline\hline
\end{tabular}
\lbltab{branchings}
\caption{Our preliminary results for the central values of the
  $\eta\pi$ branching fraction: the vector and scalar contributions
  are shown separately and compared to earlier evaluations.}
\end{table}






\section*{Aknowledgements}
We acknowledge support by in2p3-th\'eorie and by the European
Community Research Grant N$^\circ$ 283286 (HadonPhysics3)

\end{document}